\documentclass[12pt]{article}
\title{Classical solution of a sigma model in curved background}

\def\ba{\begin{array}}
\def\ea{\end{array}}
\def\be{\begin{equation}}
\def\ee{\end{equation}}

\def\eqn{equation}
\def\tfn{transformation}
\def\fn{function}
\def\-1{^{-1}}
\def\half{\frac{1}{2}}

\def\e{{\rm e}}
\def\real{{\bf R}}
\def\cd{{\cal {D}}}
\def\cg{{\cal G}}
\def\tcg{\tilde{{\cal G}}}

\def\sm{$\sigma$--model}

\def\pltd{Poisson--Lie T--dualit}
\def\dd{Drinfel'd double}

\def\email{e--mail: }
\author{Ladislav Hlavat\'y \\
Faculty of Nuclear Sciences and Physical Engineering, \\
Czech Technical University\\
B\v rehov\'a 7, 115 19 Prague 1, Czech Republic\\
\email{hlavaty@fjfi.cvut.cz}}
\begin{document}
\maketitle \abstract{We have solved equations of motion of a \sm {} in curved background using the fact that the
Poisson--Lie T-duality can transform them into the equations in the flat one. For finding solution of the flat
model we have used transformation of coordinates that makes the metric constant. The T-duality transform was
then explicitly performed.} \vskip 5mm\noindent PACS: 02.30.Ik, 04.62.+v, 11.10.Lm, 11.27.+d \vskip 5mm
\noindent Keywords: Sigma model, T--duality, Poisson--Lie T--duality, curved background, explicit solution

\section{Introduction}
Klim\v{c}\'{\i}k and \v{S}evera in their seminal work \cite{klse:dna} described  the conditions and procedure
for transforming solutions of a \sm{} to those of a dual one. Namely, let us assume that the \sm{} is defined on
a Lie group $G$ on which a covariant second order tensor field $F$ is given. The classical action of the \sm{}
then is \be S_F[\phi]=\int d^2x\, \partial_- \phi^{\mu}F_{\mu\nu}(\phi) \partial_+ \phi^\nu \label{sigm1} \ee
where the functions $ \phi^\mu:\ \real^2\rightarrow \real,\ \mu=1,2,\ldots,{\dim}\,G$ are obtained by the
composition $\phi^\mu=y^\mu\circ\phi \ $ of a map $ \phi:\real^2\rightarrow G$ and a coordinate map
$y:U_g\rightarrow \real^n,\ n={\dim}\,G$ of a neighborhood of an element $\phi(x_+,x_-)=g\in G$. If $F$
satisfies \be {\cal L}_{v_i}(F)_{\mu\nu}= F_{\mu\kappa}v^\kappa_j\tilde f_i^{jk}v^\lambda_kF_{\lambda\nu}, \
i,\mu,\nu=1,\ldots,\dim\, G \label{klimseveq}\ee where $v_i$ form a basis of left--invariant fields on $G$ and
$\tilde f_i^{jk}$ are structure coefficients of a Lie group $\tilde G, \ \dim\,\tilde G=\dim\, G$, then there is
a relation between solutions of the equations of motion for $S_F$ and $S_{\tilde F}$ where $\tilde F_{\mu\nu}$
is a second order tensor field on $\tilde G$.

The relation between the solutions $\phi(x_+,x_-)$ of the model given by $F$ and $\tilde \phi(x_+,x_-)$ of the
model given by $\tilde F$ is given by two possible decompositions of elements $d$ of Drinfel'd double

\be d=g.\tilde h= \tilde g.h \label{decofdd}\ee where $g,h\in G,\ \tilde g,\tilde h\in \tilde G$.

The map $ \tilde h:\real^2\rightarrow \tilde G$ that we need for this duality transform satisfies the equations
\cite{kli:pltd}

\be ((\partial_+ \tilde h).\tilde h\-1)_j = -A_{+,j}:= -v^\lambda_jF_{\lambda\nu}(\phi)\partial_+\phi^\nu
\label{btp}\ee\be ((\partial_- \tilde h).\tilde h\-1)_j = -A_{-,j}:=\partial_-\phi^\lambda
F_{\lambda\nu}(\phi)v^\nu_j \label{btm}\ee

Even though the equations (\ref{decofdd}--\ref{btm}) define the so called \pltd y transformation their solution
is usually very complicated to use them for finding the solutions. There are three steps in performing the
transformation:
\begin{enumerate}
\item You must know a solution $\phi(x_+,x_-)$ of the \sm{} given by $F$.
\item For the given $\phi(x_+,x_-)$ you must find $\tilde h(x_+,x_-)$ i.e. solve the system of PDE's (\ref{btp},\ref{btm}).
\item For given $d(x_+,x_-)=\phi(x_+,x_-).\tilde h(x_+,x_-)\in D$ you must find the decomposition $d(x_+,x_-)=\tilde
\phi(x_+,x_-).h(x_+,x_-)$ where $\tilde \phi(x_+,x_-)\in \tilde G,\ h(x_+,x_-)\in G$.
\end{enumerate}

The goal of this paper is to present an example of a three--dimensional \sm{} with nontrivial (i.e. curved)
background for which all the three steps can be done so that the \sm{} can be explicitly solved by this
transformation. The tensor $\tilde F$ of this model is
 \be \label{tilF}\tilde F_{\mu\nu}(\tilde y)= \left(\matrix{ - \frac{{\tilde y_1}^2}{{\kappa }^3 +
U\kappa \,\tilde y_1}
  & - \frac{\tilde y_1}{{\kappa }^2 + U\tilde y_1}  & \frac{1}
  {\kappa } \cr \frac{\tilde y_1}{{\kappa }^2 + U\tilde y_1} & \frac{\kappa }
  {{\kappa }^2 + U \tilde y_1} & 0 \cr \frac{\kappa }
  {{\kappa }^2 + U \tilde y_1} & - \frac{U}{{\kappa }^2 + U \tilde y_1}
  & 0 \cr } \right).
  \ee
where $U$ and $\kappa$ are constants. The Gauss curvature of the metric $\tilde G_{\mu\nu}(\tilde y):=(\tilde
F_{\mu\nu}(\tilde y)+ \tilde F_{\nu\mu}(\tilde y))/2$ is \be R=\frac{7\,U^4}{8\,\kappa \,{\left( {\kappa }^2 +
        U\,\tilde y_1\right) }^2}\ee so that for $U\neq 0$ we have a \sm{} in a curved background (and with nontrivial torsion). The equations of
motion have the form \begin{equation}\label{eqmot}
 \partial_- \partial_+ \phi^{\mu}+\Gamma_{\nu\lambda}^\mu \partial_- \phi^{\nu}\partial_+ \phi^{\lambda}=0
\end{equation}
where
\begin{equation}\label{chri}  \Gamma_{\nu\lambda}^\mu:=\frac{1}{2}\tilde G^{\mu\rho}(\tilde F_{\rho\lambda,\nu}
+\tilde F_{\nu\rho,\lambda}-\tilde F_{\nu\lambda,\rho}).
\end{equation}

\section{T-duality of the model}
The reason why the above given model can be solved is that it is T-dual to a model with the flat background
(Actually it was constructed in this way). It is easy to check that the tensor $\tilde F$ satisfies the
equations dual to (\ref{klimseveq}) \be {\cal L}_{\tilde v_i}(\tilde F)_{\mu\nu}= \tilde F_{\mu\kappa}\tilde
v^\kappa_j f_i^{jk}\tilde v^\lambda_k\tilde F_{\lambda\nu}, \  \label{klimseveqdual}\ee for vector fields on
$\real^3$ that are left--invariant with respect to the Abelian group structure and $f_i^{jk}$ being structure
constants of the Lie algebra given by
\begin{equation}\label{bia2} [T^1,T^2]=0,\ [T^2,T^3]=T^1,\ [T^3,T^1]=0.
  \end{equation}
It means that the equations of motion (\ref{eqmot}) of the \sm{} can be rewritten (see \cite{klse:dna},
\cite{kli:pltd}) as \eqn s on the six--dimensional Drinfel'd double $D$ -- connected Lie group whose Lie algebra
$\cal D$ admits a decomposition
\[ \cd =\tcg + \cg
\] into two subalgebras that are maximally isotropic with respect to a bilinear, symmetric, nondegenerate,
ad--invariant form.

In this case, the subalgebras $\tcg$ and $\cg$ are the three--dimensional Abelian  and the second Bianchi
algebra (\ref{bia2}) so that the \pltd y reduces to the nonabelian T-duality.
The tensor field $\tilde F$ can be obtained as
\begin{equation}\label{tilF1}
  \tilde F(\tilde\phi)=(E+\tilde\pi(\tilde\phi))\-1
\end{equation}
where
\begin{equation}\label{E0}
   E = \left(\matrix{ 0  & U&  {\kappa }
   \cr 0& {\kappa }  & 0
   \cr \kappa   &0  & 0 \cr } \right)
\end{equation}
and the matrix \fn{} $\tilde\pi(\tilde \phi)$ follows from the adjoint representation of $\tilde G$ on $\cd$
(see e.g. \cite{hlasno:3dsm2}). Similarly, the tensor field $F$ of the dual \sm{} can be obtained as
\begin{equation}\label{F}
   F(\phi)=e(\phi)E\,e(\phi)^t,
\end{equation}
where the matrix $e(\phi)$ is the vielbein field on the group $G$ corresponding to the second Bianchi algebra
(\ref{bia2}) and $e(\phi)^t$ is its transpose. From this formula one gets
\begin{equation}\label{mtz}
    F_{\mu\nu}(\phi^\rho)=\left(\matrix{ 0 & U & \kappa  \cr 0 & \kappa  & 0 \cr \kappa  & U\,\phi^2 & 2\,
   \kappa \,\phi^2 \cr  }\right)
\end{equation}
The metric of this model is flat in the sense that its Riemann tensor vanishes.

\section{Solution of the curved model}
In the following subsections we are going to perform the above given steps of the duality transform between
solutions of \eqn s of motion for $S_F$ and $S_{\tilde F}$.
\subsection{Solution of the flat model} Even though we know that the model given by the tensor (\ref{mtz}) is on
the flat background it is not easy to find the functions $\phi^\mu(x_+,x_-)$ that solve the equation of motion
given by the action $S_F[\phi]$ because the Christoffel symbols are not zero in spite of the fact that the
metric is flat. To solve the equation of motion we must express $\phi^\mu$ in terms of coordinates $\xi$ for
which the metric become constant. This was done in \cite{tur:dipl} for even more general forms of flat metrics.

Transformation of coordinates
\[ \phi^1= \xi_1  -  2\,\xi_2\,\Omega -  \frac{8\,{\Omega }^3}{3}
+ \frac{U}{4\,\kappa}(\xi_2 + 2\,{\Omega }^2)^2\]
\begin{equation}\label{flatcoors}
    \phi^2=\xi_2 + 2\,{\Omega }^2
\end{equation}
\[ \phi^3= 2\,\Omega - \frac{ U}{2\,\kappa }(\xi_2 + 2\,{\Omega }^2) \]
where $\Omega=\xi_3/2+\xi_2 U/(4\kappa)$ transform the metric obtained as the symmetric part of (\ref{mtz}) to
constant
\[ G'(\xi)=\left(\matrix{ 0 & U/2 & \kappa  \cr U/2 & \kappa  & 0 \cr \kappa  & 0 & 0\cr } \right).\]
and equations of motion transform to the wave equations so that
\begin{equation}\label{solxi}
    \xi_j(x_+,x_-)=W_j(x_+)+Y_j(x_-)
\end{equation}
with arbitrary $W_j$ and $Y_j$. Functions $\phi^\mu(x_+,x_-)$ that solve the equations of motion for $S_F[\phi]$
then follow from (\ref{flatcoors}) and (\ref{solxi}).

This finishes the first step in obtaining the solution of the \sm{} in the curved background by the duality
transform. The second step in the duality transform requires solving the system (\ref{btp},\ref{btm}).
\subsection{Solution of the system (\ref{btp},\ref{btm})} The coordinates $\tilde h_\nu$ in the
Abelian group $\tilde G$ can be chosen  so that the left--hand sides of the equations (\ref{btp},\ref{btm}) are
just $\partial_\pm \tilde h_{\nu}$. The right--hand sides are \be A_+= \left(\matrix{  U\,\partial_+\phi^2 +
\kappa \,\partial_+\phi^3\cr
  - \kappa\,\phi^3 \,\partial_+\phi^3  +
  \kappa\, \partial_+\phi^2\ - U\,\phi^3\,\partial_+\phi^2 \cr
  \kappa \,\partial_+\phi^1 + U\,\phi^2 \,\partial_+\phi^2 +
   2\,\kappa \,\phi^2\partial_+\phi^3\cr}\right) \ee
   \be A_-= \left(\matrix{ -\kappa \,\partial_-\phi^3 \cr
  -U\,\partial_-\phi^1   -
   \kappa \,\partial_-\phi^2 - U\,\,\phi^2\partial_-\phi^3 +
   \kappa\,\phi^3 \,\partial_-\phi^3\cr
  -\kappa \,\partial_-\phi^1   -
   2\,\kappa\,\phi^2 \,\partial_-\phi^3\cr}\right) \ee
and for the solution $\phi^\mu(x_+,x_-)$ found in the previous section they become rather extensive expressions
in $W(x_+)$ and $Y(x_-)$. Nevertheless, the equations (\ref{btp},\ref{btm}) can be solved and the general
solution is
\begin{eqnarray}\label{htil} \tilde h_1(x_+,x_-)&=&
   \kappa\left( \,Y_3(x_-)-
   W_3(x_+)\, \right)-
   U\,W_2(x_+)  -{U\,{\Omega }^2 },\nonumber
\\
    \tilde h_2(x_+,x_-)&=&  \kappa\left( \,Y_2(x_-)-
   W_2(x_+)\, \right)+   U\,Y_1(x_-) + \frac{U }{2}\,\beta(x_+,x_-) +\nonumber
\\&& \frac{U }{2}\,\left( W_2(x_+)\,
         Y_3(x_-) -
        W_3(x_+)\,
         Y_2(x_-) \right)+
  \frac{2\,U}{3}{\Omega }^3-\\&& \frac{U^2}{2\,\kappa }
      {\left( \half( W_2(x_+)+Y_2(x_-)) + {\Omega }^2 \right) }^2 ,\nonumber
\\ \tilde h_3(x_+,x_-)&=&\kappa\left( \,Y_1(x_-)-
   W_1(x_+)\, \right)\,+\,  \kappa\left( \,W_2(x_+)\,
    Y_3(x_-) - W_3(x_+)\,    Y_2(x_-)\, \right)+\nonumber
\\&& C\,+\,
   \kappa\,\beta(x_+,x_-)\,
  - U{\left( \half( W_2(x_+)+Y_2(x_-)) + {\Omega }^2 \right) }^2 \nonumber\end{eqnarray}
where $C$ is a constant, \[\Omega=\half( W_3(x_+)+Y_3(x_-))+\frac{U}{4\,\kappa}( W_2(x_+)+Y_2(x_-)) \] and the
function $\beta$ solves
\[ \partial_+ \beta = W_2'(x_+)\,
    W_3(x_+) -
 W_3'(x_+)\,
    W_2(x_+),\]
\[ \partial_- \beta = Y_2(x_-)\,
    Y_3'(x_-) -
 Y_3(x_-)\,
    Y_2'(x_-).\]
\subsection{Dual decomposition of elements of the \dd} The final step in the dual transformation follows from
the possibility of rewriting $$d(x_+,x_-)=\phi(x_+,x_-).\tilde h(x_+,x_-)$$ as $\tilde \phi(x_+,x_-).h(x_+,x_-)$
where $\phi(x_+,x_-),h(x_+,x_-)\in G,$ and $\tilde \phi(x_+,x_-), \tilde h(x_+,x_-)\in \tilde G$. As both $G$
and $\tilde G$ are solvable (even nilpotent) we can write all group elements as product of elements of
one--parametric subgroups and the two possible decompositions yield an equation for $\tilde \phi_\mu$ and
$h^\nu$ in terms of $\tilde h_\lambda$ and $\phi^\rho$
\begin{equation}\label{decomp1}
    \e^{\phi^1T_1}\e^{\phi^2T_2}\e^{\phi^3T_3}\e^{\tilde h_1\tilde T^1}\e^{\tilde h_2\tilde T^2}\e^{\tilde h_3\tilde T^3}
 = \e^{\tilde \phi_1\tilde T^1}\e^{\tilde \phi_2\tilde T^2}\e^{\tilde \phi_3\tilde
T^3}\e^{h^1T_1}\e^{h^2T_2}\e^{h^3T_3}.
\end{equation}
To solve it might be rather complicated in general but in this case when the only nonzero Lie products are
\[ [T_2,T_3]=T_1,\ [T_2,\tilde T^1]=-\tilde T^3,\ [T_3,\tilde T^1]=\tilde T^2 \] it can be easily done.
We can use the Baker--Campbell--Hausdorff formula that now implies \be \e^A\e^B=\e^B\e^A\e^{[A,B]}
\label{zamenag}\ee and by repeated application of this formula we get
\begin{equation}\label{soltilphi}
    \tilde \phi_1=\tilde h_1,\ \tilde \phi_2=\tilde h_2+\tilde h_1\phi^3,\
    \tilde \phi_3=\tilde h_3-\tilde h_1\phi^2
\end{equation} and $ h^\nu=\phi^\nu,\ \nu=1,2,3$.
Inserting (\ref{flatcoors}), (\ref{solxi}) and (\ref{htil}) into (\ref{soltilphi}) we get the solution of the
equations of motion for the \sm{} given by the action $S_{\tilde F}$ where $\tilde F$ is given by (\ref{tilF}).

An example of a simple solution dependent on both $x_+$ and $x_-$ is
\begin{eqnarray}\label{soltilphi1}
    \tilde \phi_1(x_+,x_-)&=&-U \sin t\,\cos x,\nonumber\\
    \tilde \phi_2(x_+,x_-)&=&-2\kappa\,\cos t \sin x+\frac{U^2}{2\kappa}\,\sin^2 t\cos^2 x,\\
    \tilde \phi_3(x_+,x_-)&=&U \sin^2 t\,\cos^2 x\nonumber
\end{eqnarray}\ where $ t=(x_++x_-)/2,\  x=(x_+-x_-)/2$ or more generally
\begin{eqnarray}\label{soltilphi2}
    \tilde \phi_1(x_+,x_-)&=&-\frac{U}{2}\,\left(Y_2(x_-)+W_2(x_+)\right),\nonumber\\
    \tilde \phi_2(x_+,x_-)&=&\kappa\,\left(Y_2(x_-)-W_2(x_+)\right)+\frac{U^2}{8\kappa}\,\left(Y_2(x_-)+W_2(x_+)\right)^2,\\
    \tilde \phi_3(x_+,x_-)&=&\frac{U}{4}\,\left(Y_2(x_-)+W_2(x_+)\right)^2\nonumber
\end{eqnarray}\
obtained from (\ref{soltilphi}) for $$Y_1=W_1=0,\ Y_3=-\frac{U}{2\kappa}Y_2,\ W_3=-\frac{U}{2\kappa}W_2,\
W_2,Y_2 \ {\rm arbitrary}.$$ Another very simple solution is
\begin{equation}\label{soltilphi3}
    \tilde \phi_1(x_+,x_-)=0,\
    \tilde \phi_2(x_+,x_-)=U\,Y_1(x_-),\
    \tilde \phi_3(x_+,x_-)=\kappa\left(Y_1(x_-)-W_1(x_+)\right).
\end{equation}obtained for $Y_2=W_2=Y_3=W_3=0,\ W_1,Y_1 \ {\rm arbitrary}
$.

\section{Conclusions}
We have explicitly solved the equations of motion of the three--dimensional \sm{} in the curved background
(\ref{tilF}) by the Poisson--Lie T-duality transformation. The solution $\tilde\phi(x_+,x_-)$ is given by
composition of the formulas (\ref{soltilphi}), (\ref{htil}), (\ref{flatcoors}) and (\ref{solxi}). Even though
the transformation is known for more than ten years it is for the first time when it was used, to the best
knowledge of the author, for finding an explicit solution. The reason may be that performing the three steps of
the transformation mentioned in the Introduction may be rather difficult in general.

To solve the equations of motion we have used the fact that we know several \sm s in the curved background that
can be transformed to the flat ones (see \cite{hlasno:3dsm2}). We were also able to find the \tfn{} of group
coordinates of the flat model to those for which the metric is constant. In the latter coordinates solution of
the flat \sm{} reduces to the solution of the wave \eqn. Performability of the next two steps of the \pltd y
transformation depends critically on the complexity of the structure of \dd{} where the \sm s live. In our case
one of the subgroup of the decomposition of the \dd{} was Abelian and the other one nilpotent. Because of that
the systems of \eqn s (\ref{btp}) and (\ref{btm}) separate and the formula (\ref{zamenag}) for solution of
(\ref{decomp1}) can be used. More complicated cases are under investigation now.

Let us note that in \cite{hlasno:pcmnsg} we have tried to solve the \eqn s of motion for \sm s on the solvable
groups with curved backgrounds by the Inverse scattering method. It turned out that for Lax pairs linear in
currents $g\-1\partial_\mu g$ the  \sm s solvable by the Inverse scattering method must have constant
Christoffel symbols which is not the case of (\ref{tilF}).

The author is grateful to Libor \v Snobl for valuable comments.

\end{document}